\documentclass[aps,prb,twocolumn,superscriptaddress,
preprintnumbers,amsmath,amssymb
,floatfix
]{revtex4}
\usepackage{graphicx}        % standard LaTeX graphics tool
\usepackage{mathrsfs} 

\usepackage{graphicx}
\usepackage{dcolumn}
\usepackage{bm}
\usepackage{amsmath}
\usepackage{amssymb}
\usepackage{revsymb}

\begin{document}

\affiliation{
Department of Physics and Astronomy, Georgia State
University, Atlanta, Georgia 30303, USA}

\title{Spaser as Nanoscale Quantum Generator and Ultrafast Amplifier}

%% Notice placement of commas and superscripts and use of &
%% in the author list

\author{Mark I. Stockman}
\affiliation{
Department of Physics and Astronomy, Georgia State
University, Atlanta, Georgia 30303, USA}
\affiliation{
Max Planck Institute for Quantum Optics,
Hans-Kopfermann-Strasse 1, 85748 Garching, Germany}
\affiliation{
Ludwig Maximilian University Munich, Am
Coulombwall 1, 85748 Garching, Germany}
\email{mstockman@gsu.edu}
\homepage{http://www.phy-astr.gsu.edu/stockman}

\date{\today}

\begin{abstract}
Nanoplasmonics has recently experienced explosive development with many
novel ideas and dramatic achievements in both fundamentals and
applications. 
%
%Among numerous applications of nanoplasmonics are
%efficient labels and sensitive assays for biomedical applications and
%defense, nanoantennas for solar cells and photodetectors, near-field
%scanning optical microscopes, concentrators of laser energy for
%thermal-assisted ultradense magnetic recording, nanoplasmonic-mediated
%thermal cancer treatment, etc. 
%
The spaser has been predicted and observed experimentally as an active
element -- generator of coherent local fields. Even greater progress
will be achieved if the spaser could function as a ultrafast
nanoamplifier -- an optical counterpart of the MOSFET
(metal-oxide-semiconductor field-effect transistor). A formidable
problem with this is that the spaser has the inherent feedback causing
quantum generation of nanolocalized surface plasmons and saturation
and consequent
elimination of the net gain, making it unsuitable for amplification. We
have overcome this inherent problem and shown that the spaser can
perform functions of an ultrafast nanoamplifier in two modes: transient
and bistable. On the basis of quantum density matrix (optical Bloch)
equations we have shown that the spaser amplifies with gain $\gtrsim 50$,
the switching time $\lesssim 100$ fs (potentially, $\sim 10$ fs). 
This prospective 
spaser technology will further broaden both fundamental and applied
horizons of nanoscience, in particular, enabling ultrafast
microprocessors working at $10-100$ THz clock speed. Other prospective
applications are in ultrasensing, ultradense and ultrafast information
storage, and biomedicine. The spasers are based on metals and, in
contrast to semiconductors, are highly resistive to ionizing radiation,
high temperatures, microwave radiation, and other adverse environments.
%
%This will allow for environmentally-robust ultrafast microprocessors, in
%particular, for nuclear industries, space research, and defense. 
\end{abstract}
\maketitle

\section{Introduction}
\label{Introduction}

Nanoplasmonic phenomena (see, e.g.,  Ref.\
\onlinecite{Novotny_Hecht_2006_Principles_of_Nanooptics})
unfold on the spatial scale between the skin
depth $l_s\approx 25$ nm (in noble metals) and the nonlocality radius
$l_{nl}\sim v_F/\omega\sim 1$ nm, where $v_F$ is the electron speed at
the Fermi surface, and $\omega$ is optical frequency. Nanoplasmonics is
ultrafast: the temporal scale
of the nanoplasmonic phenomena is between the coherent time of hundred
attoseconds defined by the inverse bandwidth of the plasmonic frequency
range (between uv and mid ir for plasmonic metals and doped
semiconductors) 
\cite{Stockman_Kling_Kleineberg_Krausz_Nature_Photonics_2007,%
Stockman_NJP_2008_Ultrafast_Nanoplasmonics_under_Coherent_Control} and
the surface plasmon (SP)
relaxation time $\gamma_p^{-1}\sim 10-100$ fs (for noble metals in the
visible to near-ir frequency range)
\cite{Bergman_Stockman:2003_PRL_spaser}.
Not just a promise anymore
\cite{Atwater_2007_Sci_Amer_Plasmonics},
nanoplasmonics has delivered a number of important applications: 
ultrasensing
\cite{Van_Duyne_Nature_Materials_2008_Biosensing_with_Plasmonic_Nanosensors},
scanning near-field optical microscopy 
\cite{Novotny_Hecht_2006_Principles_of_Nanooptics,Lewis_et_al_Laser_World_SNOMs},
SP-enhanced photodetectors 
\cite{Tang_et_al_Nat_Phot_2008_Dipole_Nanoantenna_Ge_Photodetector},
thermally assisted magnetic recording
\cite{Challener_et_al_Nati_Phot_2009_Plasmonic_Heat_Assisted_Recording},
generation of extreme uv
\cite{Kim_et_al_Nature_2008_Nanoplasmonic_EUV},
biomedical tests
\cite{Nagatani_et_al_Sci_Techn_Adv_Mater_2006_Immunochromatographic_Tests,%
Van_Duyne_Nature_Materials_2008_Biosensing_with_Plasmonic_Nanosensors},
SP-assisted thermal cancer treatment 
\cite{Halas_et_PNAS_2003_Nanoshell_NIR_Cancer_Therapy},
and many others.

To continue its vigorous development, nanoplasmonics needs an active
device -- near-field generator and amplifier of nanolocalized optical
fields, which has until recently been absent. A nanoscale amplifier in
microelectronics is the metal-oxide-semiconductor field effect transistor 
(MOSFET)
\cite{Kahng_MOSFET_Patent_1963,Tsividis_MOS_Transistor_Book_1999}, which has
enabled all contemporary digital electronics, including computers and
communications and formed the present day technology as we know it.
However, the MOSFET is limited by frequency and bandwidth to $\lesssim
100$ GHz, which is already a limiting factor in further
technological development. Another limitation of the MOSFET is its high
sensitivity to temperature, electric fields, and ionizing radiation,
which limits its use in extreme environmental conditions.

An active element of nanoplasmonics is the SPASER (Surface Plasmon
Amplification by Stimulated Emission of Radiation) that was proposed
\cite{Bergman_Stockman:2003_PRL_spaser} (see also 
Ref.\ \onlinecite{Stockman_Nat_Phot_2008_Spasers_Explained}) as a nanoscale quantum
generator of nanolocalized coherent and intense optical fields. It has
recently been observed experimentally
\cite{Noginov_et_al_Nature_2009_Spaser_Observation}. However, there is a
formidable problem to set the spaser as a nanoscale quantum amplifier. A
principal difference between a laser (quantum optical oscillator) and a
quantum optical amplifier is in that the amplifier does not possess
feedback: it does not have mirrors and any parasitic feedback due to
light scattering should be carefully minimized to prevent spontaneous
generation. In a sharp contrast, in the spaser, as we discuss below in
this Introduction, the feedback is always inherently present because the
metal plasmonic nanoparticle (spaser's core) supports SP modes whose
fields exert periodic perturbation on the gain medium causing the
feedback, which {\it principally} cannot be removed. This feedback will cause
the SP generation and the subsequent saturation of the gain. For any
spaser (or any laser, for that matter) in a stationary (CW) regime the net
amplification must be zero (which is a condition of the stable CW
operation). Therefore, one might assume that the spaser cannot serve as
a nanoamplifier. 
 
In this article we have solved this problem suggesting two approaches
to setting the spaser as a nanoamplifier: 
(i) Dynamic or transient approach,
which based on the fact that during a femtosecond transient process
after the population inversion is created but before the CW regime is
established, the spaser possesses a net amplification, and 
 (ii) Bistability
approach that is based on the inclusion of a saturable absorber in the
gain medium, which prevents spontaneous generation and sets the spaser
as a bistable (logical) ultrafast nanoamplifier. 
This is done
employing a quantum-kinetic theory that is based on density-matrix
equations, which are the adaptation of the optical Bloch equations of
the laser theory for the case of the spaser. From these equations we
have also obtained new and potentially useful results describing not
only the ultrafast dynamics of the spaser but also its CW mode:
``spasing curve'' (that is a relation between the SP population of the
spasing mode and the pumping rate) and also the linewidth of the
spaser. These results provide an excellent qualitative explanation of the recent
experimental data \cite{Noginov_et_al_Nature_2009_Spaser_Observation}.

On the basis of the present theory, one may envision femtosecond-cycle
nanoplasmonic chips with a high degree of integration where spasers
communicate and control each other through their local optical fields or
are connected with nanoplasmonic wires. These can perform ultrafast
microprocessor functions. The spasers can also be integrated with
nano-photodetectors and nanosensors to perform complex functions of
intelligent ultrafast detection and sensing. In contrast to
semiconductor technology, the spasers are based on metals and,
therefore, are highly resistive to ionizing radiation, high temperatures
and other adverse environments, with possible applications in space,
nuclear industries and defense. These functions of spaser will further
widen both the fundamental and applied horizons of nanoplasmonics and,
generally, science and technology.

\subsection{Spaser Fundamentals}
We have recently discussed the physical principles of the spasing
\cite{Stockman_Nat_Phot_2008_Spasers_Explained} but will briefly
reiterate them here for the sake of completeness and self-containment.
The spaser is a nanoplasmonic counterpart of the laser
\cite{Bergman_Stockman:2003_PRL_spaser,%
Stockman_Nat_Phot_2008_Spasers_Explained}. The laser has two principal
elements: resonator (or cavity) that supports photonic mode(s) and the
gain (or active) medium that is population-inverted and supplies energy
to the lasing mode(s). An inherent limitation of the laser is that the
size of the laser cavity in the propagation direction is at least half
wavelength and practically more than that even for the smallest lasers
developed \cite{Hill_et_al_2007_Nat_Phot_2007_Nanolasers,%
Hill_et_al_Opt_Expr_2009_Polaritonic_Nanolaser,%
Oulton_Sorger_Zentgraf_Ma_Gladden_Dai_Bartal_Zhang_Nature_2009_Nanolaser}. 
In the spaser \cite{Bergman_Stockman:2003_PRL_spaser} 
this limitation is overcome. The spasing modes are surface plasmons (SPs)
whose localization length is on the nanoscale 
\cite{Stockman:2001_PRL_Localization}
and is only limited by the minimum inhomogeneity
scale of the plasmonic metal and the nonlocality radius 
\cite{Larkin_Stockman_Nano_Letters_5_339_2005_Imperfect_Perfect_Lens}
$l_{nl}\sim 1$ nm. So, the spaser is truly nanoscopic, whose minimum
total size can be just a few nm.

\begin{figure}
\centering
\includegraphics[width=.45\textwidth]
{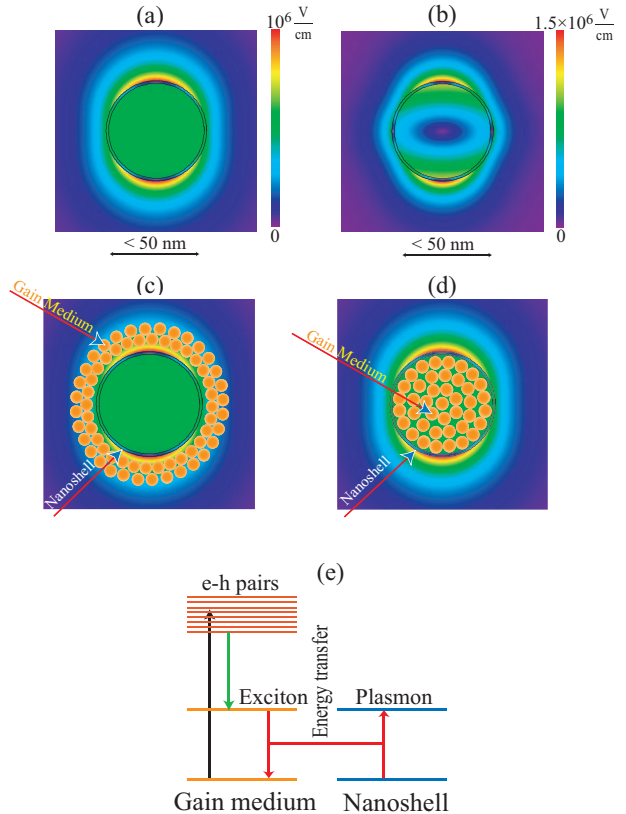}
\caption{\label{Spaser_Schematic.eps}
Schematic of spaser geometry, local fields, and fundamental processes
leading to spasing. (a) Nanoshell geometry and the local optical field
distribution for one SP in an axially-symmetric dipole mode.
The nanoshell has aspect ratio $\eta=0.95$. The local field magnitude is
color-coded by the scale bar in the right-hand side of the panel
(b) The same as (a) but for a quadrupole mode. (c) Schematic
of a nanoshell spaser where the gain medium is outside of the shell, on
the background of the dipole-mode field. (d) The same as (c) but for the
gain medium inside the shell. (e) Schematic of the spasing process. The
gain medium is excited and population-inverted
by an external source, as depicted by the black
arrow, which produces electron-hole pairs in it. These pairs relax, as
shown by the green arrow, to form the excitons. The excitons undergo
decay in the ground state emitting SPs in the nanoshell. The
plasmonic oscillations of the nanoshell stimulates this emission,
supplying the feedback for the spaser action.
}
\end{figure}
%--------------------------------------------------------------------

The resonator of a spaser can be any plasmonic metal nanoparticle whose
total size $R$ is much less than the wavelength $\lambda$ and whose metal
thickness is between $l_{nl}$ and $l_s$, which supports a SP mode with required
frequency $\omega_n$. This
metal nanoparticle should be surrounded by the gain medium that overlaps
with the spasing SP eigenmode spatially and whose emission line overlaps
with this eigenmode spectrally
\cite{Bergman_Stockman:2003_PRL_spaser}. 
As an example, we consider a model of a nanoshell spaser
\cite{Gordon_Ziolkowski_Ope_Express_2007_Nanoparticle_Laser,%
Stockman_Nat_Phot_2008_Spasers_Explained},
which is illustrated in Fig.\ \ref{Spaser_Schematic.eps}. Panel (a)
shows a silver nanoshell carrying a single SP (plasmon population number 
$N_n=1$) in the dipole
eigenmode. It is characterized by a uniform field inside the core and
hot spots at the poles outside the shell with the maximum field reaching
$\sim 10^6$ V/cm. Similarly, Fig.\ \ref{Spaser_Schematic.eps} (b) shows
the quadrupole mode in the same nanoshell. In this case, the mode electric field is
non-uniform, exhibiting hot spots of $\sim 1.5\times 10^6$ V/cm
of the modal electric field at the poles. These high values of
the modal fields is the underlying physical
reason for very strong feedback in the spaser.
%
%In this article, we are interested in true spasers whose maximum size is
%much less than the wavelength $\lambda$ and and the maximum metal
%thickness is less than the skin depth $l_s\approx 25$ nm. This is in
%contrast to nanolasers based on surface plasmon polaritons (SPPs)
%\cite{Hill_et_al_Opt_Expr_2009_Polaritonic_Nanolaser,%
%Oulton_Sorger_Zentgraf_Ma_Gladden_Dai_Bartal_Zhang_Nature_2009_Nanolaser} 
%whose transverse modal size in on the nanoscale but the longitudinal
%size in the direction of propagation is on the order of the wavelength,
%i.e., microscopic. 
Under our conditions, the electromagnetic retardation
within the spaser volume can be safely neglected. Also, the radiation of
such a spaser is a weak effect: the decay rate of plasmonic eigenmodes
is dominated by the internal loss in the metal. Therefore, it is
sufficient to consider only quasistatic eigenmodes
\cite{Bergman_Stroud_1992,Stockman:2001_PRL_Localization} and not their
full electrodynamic counterparts
\cite{Bergman_Stroud_Phys_Rev_B_22_3527_1980_Spectral_Theory_Full_Electrodynamics},
which also is necessary for the ultrafast operation of the spaser as a
nanoamplifier.

Note that for high aspect ratios (say, $\eta=0.95$ considered
in some of the examples) and relatively small radii ($R=12$ nm) the
thickness  of the metal shell may be rather small (less than 1 nm),
which may lead to nonlocal effects including Landau damping. In this
case, the radius of the spaser can be safely increased by a factor of 2
to 3 (this will not adversely effect the spasing as the spaser is
fully scalable, as we show below in Sec.\ \ref{Stationary_Spaser}). Also
one should keep in mind that the nonlocal effects in thin nanoshells are
small due to the special symmetry reasons
\cite{Kirakosyan_Stockman_Shahbazyan_ArXiv_2009_Nanoshells}. We also
consider a number of examples with significantly higher aspect ratios
where the metal thickness is a few nanometers or larger.

For the sake of numerical illustrations of our theory, we will use the
dipole eigenmode [Fig.\ \ref{Spaser_Schematic.eps} (a)]. There are two
basic ways to place the gain medium: (i) outside the nanoshell, as shown
in panel (c), and (ii) in the core, as in panel (d), which was
originally proposed in Ref.\
\onlinecite{Gordon_Ziolkowski_Ope_Express_2007_Nanoparticle_Laser}. As we
have verified, these two designs lead to comparable characteristics of
the spaser. However, the placement of the gain medium inside the core
illustrated in Fig.\ \ref{Spaser_Schematic.eps} (d) has a
significant advantage because the hot spots of the local field are not
covered by the gain medium and are sterically available for
applications. 

Note that any $l$-multipole mode of a spherical particle
is, indeed, $2l+1$-times degenerate. This may make the spasing mode to
be polarization unstable, like in lasers without polarizing
elements. In reality, the polarization may be pinned and become stable due
to deviations from the perfect spherical symmetry, which exist naturally
or can be introduced deliberately. More practical shape for a spaser may
be a nanorod, which has a mode with the stable polarization along the
major axis.
However, a nanorod is a more complicated geometry for theoretical
treatment, and we will consider it elsewhere.

The level diagram of the spaser gain medium and the plasmonic metal
nanoparticle is displayed in Fig.\ \ref{Spaser_Schematic.eps} (e) along
with a schematic of the relevant energy transitions in the system. The
gain medium chromophores may be semiconductor
nanocrystals \cite{Bergman_Stockman:2003_PRL_spaser,%
Plum_Fedotov_Kuo_Tsai_Zheludev_Opt_Expr_2009_Toward_Lasing_Spaser}, dye
molecules \cite{Seidel_Grafstroem_Eng_Phys_Rev_Lett_94_177401_2005,%
Noginov_et_al_PRL_2008_SPP_Stimulated_Emision}, rare-earth
ions \cite{Gordon_Ziolkowski_Ope_Express_2007_Nanoparticle_Laser}, or
electron-hole excitations of a bulk semiconductor 
\cite{Hill_et_al_2007_Nat_Phot_2007_Nanolasers,%
Hill_et_al_Opt_Expr_2009_Polaritonic_Nanolaser}. For certainty, we will
use a semiconductor-science language of electrons and holes. 

The pump
excites electron-hole pairs in the chromophores [Fig.\
\ref{Spaser_Schematic.eps} (e)], as indicated by the vertical black
arrow, which relax to form excitons. The excitons constitute the two-%
level systems that are the donors of energy for the SP emission into the
spasing mode. In vacuum, the excitons would recombine emitting photons.
However, in the spaser geometry, the photoemission is strongly quenched
due to the resonance energy transfer to the SP modes, as indicated by
the red arrows in the panel. The plasmons in the spaser mode create the
high local fields that excite the gain medium and stimulate more
emission to this mode, which is the feedback mechanism. If this feedback
is strong enough and the life time of the spaser SP mode is long enough,
then an instability develops leading to the avalanche of the SP emission
in the spasing mode and spontaneous symmetry breaking, establishing the
phase coherence of the spasing state. Thus the establishment of spasing
can be called a nonequilibrium phase transition, as in the physics of
lasing.

\subsection{Brief Overview of Latest Progress in Spasers and Nanolasers}
After the original theoretical proposal and prediction of the spaser
\cite{Bergman_Stockman:2003_PRL_spaser}, there has been an active
development in this field, both theoretical and experimental. We comment
below only on some representative
publications. Among theoretical developments, a nanolens spaser has been
proposed
\cite{Li_Li_Stockman_Bergman_PRB_71_115409_2005_Nanolens_Spaser}, which
possesses a nanofocus (``the hottest spot'') of the local fields. In Refs.\
\onlinecite{Bergman_Stockman:2003_PRL_spaser,%
Li_Li_Stockman_Bergman_PRB_71_115409_2005_Nanolens_Spaser}, only the
necessary condition of spasing has been established on the basis of the
perturbation theory. 

There have been theories published describing the spaser
(or, ``nanolaser'' as sometimes it is called) phenomenologically,
on the basis of classic linear
electrodynamics by considering the gain medium as a dielectric with a
negative imaginary part of the permittivity 
\cite{Zhang_et_al_Opt_Expr_2008_NIM_Transmission_Ampification,%
Wegener_et_al_Opt_Expr_2008_Spaser_Toy_Model,%
Gordon_Ziolkowski_Ope_Express_2007_Nanoparticle_Laser,%
Wegener_et_al_PRB_2009_Metamaterials_with_Gain}. Such
electrodynamic approaches do not take into account the nature of the
spasing as a spontaneous symmetry breaking. 
This leads to principal differences of their results from the present
microscopic quantum-mechanical theory in the
region of spasing, as we discuss below in
Sec.\ \ref{CW_Kinetics} in conjunction with Fig.\
\ref{Spaser_Threshold_Linewidth_Spectra.eps}. Note that before the
spasing threshold, the above-mentioned phenomenological theories are,
in principle, applicable. 
There has also been a theoretical publication on a bowtie spaser
(nanolaser) with electrical pumping
\cite{Chang_Ni_Chuang_Opt_Expr_2008_Bowtie_Nanolaser_Theory}. It is
based on balance equations and only the CW spasing generation intensity
is described. Yet another theoretical development has been
a proposal of the lasing
spaser \cite{Zheludev_et_al_Nat_Phot_2008_Lasing_Spaser},
which is made of a plane array of spasers. The theoretical publications
mentioned above in this paragraph deal with the CW regime. 
In contrast, in this article we are most interested in the
ultrafast behavior of the spaser. 

There have also been a theoretical proposal of a spaser (``nanolaser'')
consisting of a metal nanoparticle coupled to a single chromophore
\cite{Protsenko_et_al_Phys_Rev_A71_063812_2005_Dipole_Nanolaser}. In
this paper, a dipole-dipole interaction is illegitimately used at very
small distances $r$ where it has a singularity (diverging for $r\to 0$),
leading to a dramatically overestimated coupling with the SP mode. As a
result, a completely unphysical prediction of CW spasing due to single
chromophore has been obtained
\cite{Protsenko_et_al_Phys_Rev_A71_063812_2005_Dipole_Nanolaser}. In
contrast, our theory is based on the full (exact) field of the spasing SP mode.
As our results of Sec.\ \ref{CW_Spaser} below show, hundreds of
chromophores per metal nanoparticle are realistically requited for the
spasing even under the most favorable conditions.

%There have also been 
%theoretical proposals of magnetic-resonance nanolasers (which can be
%called magnetic spasers).%
%\cite{Sarychev_Tartakovsky_PRB_2007_Nanolaser,%
%Zhu_et_al_APL_2009_Nanolaser_with_Two_Magnetic_Modes} These are
%principally different from the spaser because they explicitly depend on
%speed of light $c$ (the spaser is quasistatic and its theory does not
%contain $c$). Importantly, this distinction also leads to
%much smaller modal volume and much
%stronger feedback in the spaser (much greater values of the Purcell
%factor\cite{Purcell_PR_1946_Purcell_Factor}),
%as discussed below in conjunction with Eq.\ (\ref{qualitative}).

There has been a vigorous experimental quest toward the spaser. 
Stimulated emission of SPPs has been observed in a
proof-of-principle experiment using pumped dye molecules as an active
(gain) medium \cite{Seidel_Grafstroem_Eng_Phys_Rev_Lett_94_177401_2005}.
There have also been later
experiments that demonstrated strong stimulated emission compensating a
significant part of the SPP loss 
\cite{Noginov_et_al_PRL_2008_SPP_Stimulated_Emision,%
Zhang_et_al_Nano_Lett_2009_SPP_Stimulated_Emission,%
Zhou_Su_Peng_Zhou_Opt_Expr_2008_SPP_Amplification_in_Ag_Nanowires,%
Noginov_et_al_Opt_Expr_2008_SPP_Loss_Compensation}. As a step toward the
lasing spaser, the first experimental demonstration has been reported of
compensating Joule losses in a metallic photonic metamaterial using
optically pumped PbS semiconductor quantum dots
\cite{Plum_Fedotov_Kuo_Tsai_Zheludev_Opt_Expr_2009_Toward_Lasing_Spaser}.
In another development, a tunable free-electron light nanosource 
("light well") has been demonstrated 
\cite{Zheludev_et_el_PRL_2009_Light_Well}, which is based on spontaneous 
emission in a periodically layered metal-dielectric structure. 
There have also been
experimental investigations reporting the stimulated emission effects of
SPs in plasmonic metal nanoparticles surrounded by gain media with dye molecules 
\cite{Noginov_et_al_APB_2007_Loss_Compensation,%
Noginov_J_Nanophot_2008_Gain_Loss_Compensation}.
An electrically-pumped nanolaser with semiconductor gain medium has been
demonstrated \cite{Hill_et_al_Opt_Expr_2009_Polaritonic_Nanolaser} where the
lasing modes are SPPs. 
A nanolaser with optically-pumped semiconductor gain medium and a hybrid
semiconductor-silver SPP waveguide has been demonstrated with an
extremely tight transverse mode confinement 
\cite{Oulton_Sorger_Zentgraf_Ma_Gladden_Dai_Bartal_Zhang_Nature_2009_Nanolaser}.

Finally, an observation has been reported of a true spaser by
M.\ Noginov et al.\ 
\cite{%Noginov_at_al_OSA_META_2008_Spaser_Observation,%
%Noginov_NanoMeta_2009_Spaser_Observation,%
Noginov_et_al_Nature_2009_Spaser_Observation}.
This spaser is a chemically synthesized gold nanosphere of radius 
7 nm surrounded by a
dielectric shell of a 21 nm outer radius containing immobilized dye
molecules, which is sometimes referred to as ``Cornell nanodot''.
Under nanosecond optical pumping in the absorption band of the dye,
this spaser develops a relatively narrow-spectrum and intense visible emission
that exhibits a pronounced threshold in pumping intensity.
The observed characteristics of this spaser are
in an excellent qualitative agreement and can be fully understood
on the basis of the corresponding theoretical results 
obtained below in Sec.\ \ref{CW_Kinetics}.

This article is organized as the following. In Sec.\ \ref{DME} for the
case of the spaser, we 
formulate quantum density matrix equations (also called in the
literature the optical Bloch equations). In Sec.\
\ref{Stationary_Spaser} we consider these equations for a stationary
(CW) case where we find the conditions of spasing, the population of
the SPs in the spasing mode, and the spasing line shape. In Sec.\
\ref{CW_Kinetics} we illustrate the spasing curve and spectral
composition of the spaser generation. In Sec.\ \ref{CW_Kinetics} we
describe the CW spaser as a bistable nanodevice. 
The major results of this articles
regarding the ultrafast kinetics of the spaser as a generator and
quantum nanoamplifier are presented in Sec.\ \ref{Ultrafast}. We
briefly discuss the obtained results and conclude in Sec.\
\ref{Conclusions}.

\section{Equations of Spaser}
\label{Methods}

\subsection{Quantum Density Matrix (Optical Bloch) Equations for Spaser}
\label{DME}

The SP eigenmodes $\varphi_n(\mathbf r)$ are described by a wave
equation (with homogeneous boundary conditions)%
\cite{Stockman:2001_PRL_Localization,Bergman_Stockman:2003_PRL_spaser}
\begin{equation}
\nabla\Theta(\mathbf r)\nabla\varphi_n(\mathbf r)=
s_n\nabla^2 \varphi_n(\mathbf r)~,
\label{phin}
\end{equation}
where $n$ is the mode number, $s_n$ is corresponding eigenvalue, 
and $\Theta(\mathbf r)$ is the
characteristic function equal to 1 for $\mathbf r$ in the metal component 
and 0 in the dielectric. Note that the eigenvalues $s_n$ are all real
and contained in the range $1\ge s_n\ge 0$. The eigenmodes are
normalized by an integral over the volume $V$ of the system, 
$\int_V \left|\nabla\varphi_n(\mathbf r)\right|^2 d^3r=1$.
The physical frequency $\omega_n$ of the SPs is defined by 
an equation $\mathrm{Re}\left[s(\omega_n)\right]=s_n$, where
$s(\omega)=
\varepsilon_d/\left[\varepsilon_d-\varepsilon_m(\omega)\right]$ 
is Bergman's spectral parameter, 
$\varepsilon_d$ is the permittivity of the ambient dielectric, and
$\varepsilon_m(\omega)$ is the metal permittivity.

The electric field operator \footnote{Note that we have corrected 
a misprint in Ref.\ 
\onlinecite{Bergman_Stockman:2003_PRL_spaser} replacing the coefficient
$2\pi$ by $4\pi$.}
of the quantized SPs is%
\cite{Bergman_Stockman:2003_PRL_spaser}
\begin{equation}
\mathbf E(\mathbf r)=-
\sum_n A_n
\nabla\varphi_n(\mathbf r)(\hat a_n+\hat a_n^\dag)~,~~~ 
A_n=\left(\frac{4\pi \hbar s_n}{\varepsilon_d s^\prime_n}\right)^{1/2}~,
\label{E}
\end{equation}
where $\hat a_n^\dag$ and $\hat a_n$ are the SP creation and annihilation
operators, and $s^\prime_n=\mathrm{Re}\left[d
s(\omega_n)/d\omega_n\right]$.

The spaser Hamiltonian has the form
\begin{equation}
H=H_g+\hbar \sum_n\omega_n \hat a_n^\dag \hat a_n -
\sum_p \mathbf E(\mathbf r_p)\mathbf d^{(p)}~,
\label{H}
\end{equation}
where $H_g$ is the Hamiltonian of the gain medium, $p$ is an index (label) of
a gain medium chromophore, $\mathbf r_p$ is its coordinate vector, 
and $\mathbf d^{(p)}$ is its dipole moment operator. In this paper, we
will treat the active medium quantum mechanically but the SPs
quasiclassically, considering $\hat a_n$ as a classical quantity 
(c-number) $a_n$  with time dependence as  $a_n=a_{0n} \exp(-i\omega
t)$, where $a_{0n}$ is a slowly-varying amplitude. The number of
coherent SPs per spasing mode is then given by
$N_p=\left|a_{0n}\right|^2$. This approximation neglects the 
quantum fluctuations of the SP amplitudes. However, 
when necessary, we will take into account these quantum fluctuations,
in particular, to describe the spectrum of the spaser. 

Introducing $\rho^{(p)}$ as the density matrix of a $p$th chromophore,
we can find its equation of motion in a conventional way by commutating it
with the Hamiltonian (\ref{H}) as $i\hbar\dot\rho^{(p)}=[\rho^{(p)},H]$,
where the dot denotes temporal derivative. We will use the standard
rotating wave approximation (RWA), which only takes into account the resonant
interaction between the optical field and chromophores. We denote
$\left|1\right\rangle$ and $\left|2\right\rangle$ as the ground and
excited states of a chromophore, with the transition 
$\left|2\right\rangle \rightleftharpoons \left|1\right\rangle$
resonant to the spasing plasmon mode $n$.
In this approximation, the time dependence of the nondiagonal elements
of the density matrix is $\left(\rho^{(p)}\right)_{12}=
\bar\rho^{(p)}_{12}\exp(i\omega t)$, and $\left(\rho^{(p)}\right)_{21}=
\bar\rho^{(p)\ast}_{12}\exp(-i\omega t)$, where $\bar\rho^{(p)}_{12}$ is
a time-independent amplitude defining the coherence
(polarization) for the 
$\left|2\right\rangle \rightleftharpoons \left|1\right\rangle$
spasing transition in a $p$th chromophore of the gain medium.

Introducing a rate constant $\Gamma_{12}$ to describe the polarization
relaxation and a difference $n^{(p)}_{21}=\rho^{(p)}_{22}-\rho^{(p)}_{11}$  as 
the population inversion on this spasing transition, we derive an equation 
of motion for the non-diagonal element of the density matrix as
\begin{equation}
\dot{\bar\rho}^{(p)}_{12}=
-\left[i\left(\omega-\omega_{12}\right)+\Gamma_{12}\right]\bar\rho^{(p)}_{12}
+i n^{(p)}_{21} \Omega^{(a)\ast}_{12}~,
\label{rho12}
\end{equation}
where $\Omega^{(p)}_{12}=-A_n \mathbf d^{(p)}_{12}
\nabla\varphi_n(\mathbf r_p) a_{0n}/\hbar$ is the
Rabi frequency for the spasing transition in a $p$th chromophore, and 
$\mathbf d^{(p)}_{12}$ is the corresponding transitional dipole element.
Note that always $\mathbf d^{(p)}_{12}$ is either real or can be made real 
by a proper choice of the quantum state phases, making the
Rabi frequency $\Omega^{(p)}_{12}$ also a real quantity. 

An equation of motion for $n^{p}_{21}$ can be found in a
standard way by commutating it with $H$. To provide conditions for the
population inversion ($n^{p}_{21}>0$), we imply existence of a
third level. For simplicity, we assume that it very rapidly decays into 
the excited state $\left|2\right\rangle$ of the chromophore, so
its own populations is negligible. It is pumped by an external source from
the ground state (optically or electrically) with some rate that we will 
denote $g$. In this way, we obtain the following equation of motion:
\begin{equation}
\dot{\bar n}^{(p)}_{21}=-
4\mathrm{Im}\left[\bar\rho^{(p)}_{12}\Omega^{(p)}_{21}\right]-
\gamma_2\left(1+n^{(p)}_{21}\right)+g\left(1-n^{(p)}_{21}\right)~,
\label{n21}
\end{equation}
where $\gamma_2$ is the decay rate 
$\left|2\right\rangle \rightarrow\left|1\right\rangle$.

The stimulated emission of the SPs is described as their
excitation by the coherent polarization of the gain medium. The
corresponding equation of motion can be obtained using Hamiltonian
(\ref{H}) and adding the SP relaxation with a rate of $\gamma_n$ as
\begin{equation}
\dot a_{0n}=
\left[i\left(\omega-\omega_n\right)-\gamma_n\right]a_{0n}+
i\sum_p \rho^{(p)\ast}_{12}\Omega^{(p)_{12}}~.
\label{a0n}
\end{equation}

Another relevant process is spontaneous emission of SPs by a
chromophore into a spasing SP mode. 
 The corresponding rate $\gamma_2^{(p)}$ for a chromophore at a point
$\mathbf r_p$ can be found in a
standard way using the quantized field (\ref{E}) as
\begin{equation}
\gamma_2^{(p)}=
2\frac{A_n^2}{\hbar\gamma_n}
\left|\mathbf d_{12} \nabla\varphi_n(\mathbf r_p)\right|^2
\frac{\left(\Gamma_{12}+\gamma_n\right)^2}%
{\left(\omega_{12}-%
\omega_n\right)^2+\left(\Gamma_{12}+\gamma_n\right)^2}~.
\label{gamma_2p}
\end{equation}
As in Schawlow-Towns theory of laser-line width, this spontaneous
emission of SPs leads to the diffusion of the phase of the spasing
state. This defines width $\gamma_s$ of the spasing line as
\begin{equation}
\gamma_s=\frac{\sum_p \left(1+n^{(p)}_{21}\right)\gamma_2^{(p)}}{2(2N_p+1)}~.
\label{gamma_s}
\end{equation}
This width is small for a case of developed spasing when $N_p\gg 1$.
However, for $N_p\sim 1$, the predicted width may be too high because
the spectral diffusion theory assumes that $\gamma_s\lesssim \gamma_n$.
To take into account this limitation in a simple way, we will
interpolate to find the resulting spectral width $\Gamma_s$ of the
spasing line as $\Gamma_s=\left(\gamma_n^{-2}+\gamma_s^{-2}\right)^{-
1/2}$.

We will also examine the spaser as a bistable (logical) amplifier. 
One of the ways
to set the spaser in such a mode is to add a saturable absorber. This is
described by the same Eqs. (\ref{rho12})-(\ref{a0n}) where the
chromophores belonging to the absorber are not pumped by the external
source directly, i.e., for  them in Eq.\ (\ref{n21}) one has to set
$g=0$. 

Numerical examples are given for a silver nanoshell where the core and
the external dielectric have the same permittivity of $\varepsilon_d=2$;
the permittivity of silver is adopted from Ref.\
\onlinecite{Johnson:1972_Silver}. The following realistic parameters of the gain
medium are used (unless indicated otherwise): $d_{12}=1.5\times 10^{-
17}$ esu, $\hbar \Gamma_{12}=10$ meV, $\gamma_2=4\times
10^{12}~\mathrm{s^{-1}}$ (this value takes into account the spontaneous
decay into SPs), and density of the gain medium chromophores is
$\rho=2.4\times 10^{20}~\mathrm{cm^{-3}}$, which is realistic for dye
molecules but may be somewhat high for semiconductor quantum dots that
were proposed as the chromophores\cite{Bergman_Stockman:2003_PRL_spaser}
and used in experiments%
\cite{Plum_Fedotov_Kuo_Tsai_Zheludev_Opt_Expr_2009_Toward_Lasing_Spaser}.
We will assume a dipole SP mode and chromophores situated in the core of
the nanoshell as shown in Fig.\ \ref{Spaser_Schematic.eps} (d). This
configuration are of advantage both functionally (because the region of
the high local fields outside the shell is accessible for various
applications) and computationally (the uniformity of the modal fields
makes the summation of the chromophores trivial, thus greatly
facilitating numerical procedures).

\subsection{Equations for Stationary (CW) Regime}
\label{Stationary_Spaser}

Physically, the spaser action is a result of spontaneous symmetry
breaking when the phase of the coherent SP field is established from the
spontaneous noise. Mathematically, the spaser is described by homogeneous
differential Eqs.\ (\ref{rho12})-(\ref{a0n}).  
These equations become homogeneous
algebraic equations for the stationary [or continuous 
wave (CW)] case. 
They always have a trivial,
zero solution. However, when their determinant vanishes, they
also possess a nontrivial solution describing spasing, whose condition is
\begin{eqnarray}
&&\left(\omega_s-\omega_n+i\gamma_n\right)^{-1}\times
\\ \nonumber
&& \left(\omega_s-\omega_{21}+i\Gamma_{12}\right)^{-1}
\sum_p \left|\tilde\Omega^{(p)}_{12}\right|^2 n^{(p)}_{21}
=-1~,
\label{spasing}
\end{eqnarray}
where %$\omega_s$ is the spasing frequency, 
$\tilde\Omega^{(p)}_{12}=
-A_n \mathbf d^{(p)}_{12}
\nabla\varphi_n(\mathbf r_p)/\hbar$ is the single-plasmon Rabi frequency.
% 
%$\mathbf d^{(p)}_{12}$ is the transition dipole moment of a $p$th
%chromophore, $\varphi_n(\mathbf r_p)$ is the electric potential of the
%spasing mode at the position this chromophore, $\gamma_n$ is the decay
%rate of the SP mode,\cite{Bergman_Stockman:2003_PRL_spaser}
% $\Gamma_{12}$ is the width and $\omega_{21}$ is the 
%frequency of the spasing transition in the chromophores. Here $A_n$ is the
%amplitude of the quantized SP
%field\cite{Bergman_Stockman:2003_PRL_spaser}:
%$A_n=\left(4\pi \hbar s_n/\varepsilon_d s^\prime_n\right)^{1/2}$,
%$s_n=s(\omega_n)$, $s^\prime_n=\partial s(\omega_n)/\partial\omega_n$,
%where $s(\omega)=\left[1-\varepsilon_m(\omega)/\varepsilon_d\right]^{-1}$ is
%Bergman's spectral parameter with $\varepsilon_m(\omega)$ as the
%permittivity of the metal and $\varepsilon_d$ that of the surrounding
%dielectric. 
The population inversion of a $p$th chromophore $n^{(p)}_{21}$ is explicitly
expressed as
\begin{eqnarray}
&&n^{(p)}_{21}=\left(g-\gamma_2\right)\times
\\ \nonumber
&&\left\{g+\gamma_2+4 N_n \left|\tilde \Omega^{(p)}_{12}\right|^2\left/%
\left[\left(\omega_s-\omega_{21}\right)^2+%
\Gamma_{12}^2\right]\right.\right\}^{-1}~.
\label{n21p}
\end{eqnarray}
%where $\Omega^{(p)}_{12}=\tilde\Omega^{(p)}_{12}\sqrt{N_n}$ 
%is the Rabi frequency in a state containing $N_n$ of SP quanta.
%, $g$
%is the pumping rate per a chromophore, and $\gamma_2$ is the decay rate
%of the upper spasing level population.

From the imaginary part of Eq.\ (\ref{spasing}) we immediately find the
spasing frequency
\begin{equation}
\omega_s=\left(\gamma_n \omega_{21}+\Gamma_{12} \omega_n\right)
\left/\left(\gamma_n+\Gamma_{12}\right)\right.~,
\label{omega_s}
\end{equation}
which generally does not coincide with either the gain transition
frequency $\omega_{21}$ or the SP frequency $\omega_n$, but is between
them  (this is a frequency walk-off phenomenon similar to that of laser physics).
Substituting Eq.\ (\ref{omega_s}) back to 
Eqs.\ (\ref{spasing})-(\ref{n21p}), we obtain a system of equations
\begin{eqnarray}
&& 
\frac{\left(\gamma_n+\Gamma_{12}\right)^2}%
{\gamma_n\Gamma_{12}%
\left[\left(\omega_{21}-\omega_n\right)^2+\left(\Gamma_{12}+\gamma_n\right)^2\right]}\times%
\nonumber \\
&&\sum_p \left|\tilde\Omega^{(p)}_{12}\right|^2 n^{(p)}_{21}=1~,
\label{spasing1} \\ 
&&
n^{(p)}_{21}=\left(g-\gamma_2\right)\times%
\nonumber \\ 
&&\left[g+\gamma_2+%
\frac{4 N_n \left|\tilde\Omega^{(p)}_{12}\right|^2%
\left(\Gamma_{12}+\gamma_n\right)}%
{\left(\omega_{12}-\omega_n\right)^2+%
\left(\Gamma_{12}+\gamma_n\right)^2}\right]^{-1}~.
\label{n21p_1}
\end{eqnarray}
This system defines the
stationary (CW) number of SPs per spasing mode $N_n$.
% which enters the equations via the Rabi frequency $\Omega^{(p)}_{12}$.

Since $n^{(p)}_{21} \le 1$, from Eqs.\ (\ref{spasing1}), (\ref{n21p_1}) we immediately
obtain a necessary condition of the existence of spasing,
\begin{equation}
\frac{\left(\gamma_n+\Gamma_{12}\right)^2}%
{\gamma_n\Gamma_{12}%
\left[\left(\omega_{21}-\omega_n\right)^2+\left(\Gamma_{12}+\gamma_n\right)^2\right]}
\sum_p \left|\tilde\Omega^{(p)}_{12}\right|^2 \ge 1~.
\label{criterion}
\end{equation}
This expression is fully consistent with Ref.\
\onlinecite{Bergman_Stockman:2003_PRL_spaser}. The following order of magnitude
estimate of this spasing condition has a transparent physical meaning
and is of heuristic value,
\begin{equation}
\frac{d_{12}^2 Q N_c}{\hbar\Gamma_{12}V_n}\gtrsim 1~,
\label{qualitative}
\end{equation}
where $Q=\omega/\gamma_n$ is the quality factor of SPs, 
$V_n$ is the volume of the spasing SP mode, 
and $N_c$ is the of number of the gain medium chromophores within
this volume. Deriving this estimate, we
have neglected the detuning, i.e., set $\omega_{21}-\omega_n=0$.
We also used the definitions of $A_n$ of Eq.\ (\ref{E}) 
and $\tilde\Omega_{12}^{(p)}$ given after Eq.\ (\ref{spasing}), 
and the estimate $\left|\nabla\varphi_n(\mathbf r) \right|^2 
\sim 1/V$ following from the normalization of the SP eigenmodes 
$\int \left|\nabla\varphi_n(\mathbf r) \right|^2 d^3 r= 1$ of Ref.\
\onlinecite{Stockman:2001_PRL_Localization}. The result of Eq.\ 
(\ref{qualitative}) is, indeed, in agreement with Ref.\ 
\onlinecite{Bergman_Stockman:2003_PRL_spaser} where it was obtained in
slightly different notations.

It follows from Eq.\ (\ref{qualitative}) that for the existence of
spasing it is beneficial to have a high quality factor $Q$, a high
density of the chromophores, and a large transition dipole (oscillator
strength) of the chromophore transition. The small modal volume $V_n$
(at a given number of the chromophores $N_c$) is beneficial for this
spasing condition: physically, it implies strong feedback in the spaser.
Note that for the given density of the chromophores $\rho_c=N_c/V_n$,
this spasing condition does not explicitly depend on the spaser size,
which opens up a possibility of spasers of a very small size limited
from the bottom by only the nonlocality radius $l_{nl}\sim 1$ nm.
Another important property of Eq.\ (\ref{qualitative}) is that it
implies the quantum-mechanical nature of spasing and spaser
amplification: this condition essentially contains the Planck constant
$\hbar$ and, thus, does not have a classical counterpart. Note that in
contrast to lasers, the spaser theory and Eq.\ (\ref{qualitative}) in
particular do not contain speed of light, i.e., they are
quasistatic.

\section{Spaser in Mode of Continuous Wave Nanoscale Quantum Generator}
\label{CW_Spaser}

\subsection{Kinetics of CW Spaser}
\label{CW_Kinetics}

\begin{figure}
\centering
\includegraphics[width=.45\textwidth]
{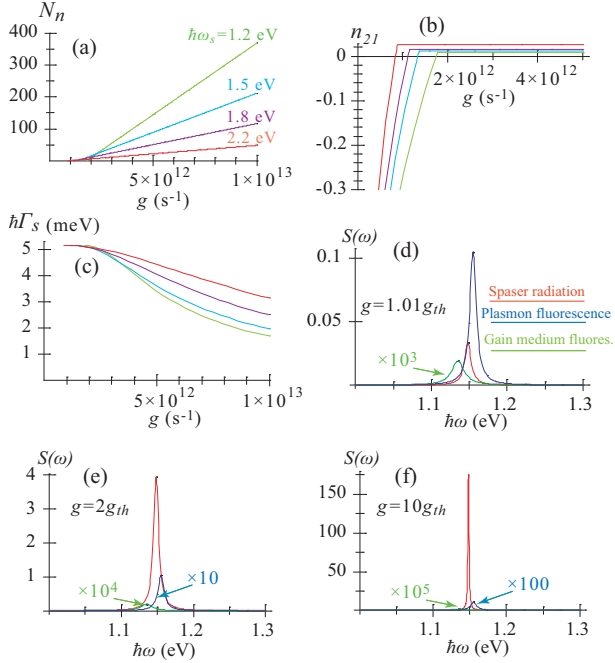}
\caption{\label{Spaser_Threshold_Linewidth_Spectra.eps}
Spaser SP population and spectral characteristics in the stationary
state. The computations are done for a silver nanoshell with the
external radius $R_2=12$ nm; the detuning of the gain medium from the
spasing SP mode is $\hbar\left(\omega_{21}-\omega_n\right)=-0.02$ eV.
The other parameters are indicated in Sec.\ \ref{Methods}. (a) Number $N_n$ of
plasmons per spasing mode as a function of the excitation rate $g$ (per
one chromophore of the gain medium). Computations are done for the
dipole eigenmode with the spasing frequencies $\omega_s$ as indicated,
which were chosen by the corresponding adjustment of the nanoshell
aspect ratio. (b) population inversion $n_{12}$ as a function of the
pumping rate $g$. The color coding of the lines is the same as in panel
(a). (c) The spectral width $\Gamma_s$ of the spasing line (expressed as
$\hbar\Gamma_s$ in meV) as a function of the pumping rate $g$. The color
coding of the lines is the same as in panel (a). (d)-(f) Spectra of the
spaser for the pumping rates $g$ expressed in the units of the threshold
rate $g_{th}$, as indicated in the panels. The curves are color coded and
scaled as indicated. 
}
\end{figure}

The ``spasing curve'', i.e., the dependence of the coherent SP population
$N_n$ on the excitation rate $g$, obtained by solving Eqs.\
(\ref{spasing1}), (\ref{n21p_1}), is shown in Fig.\
\ref{Spaser_Threshold_Linewidth_Spectra.eps} (a) for four types of the
silver nanoshells with the frequencies of the spasing dipole modes as
indicated, which are in the range from near ir ($\hbar\omega_s=1.2$ eV)
to mid visible ($\hbar\omega_s=2.2$ eV). In all cases, there is a
pronounced threshold of the spasing at an excitation rate $g_{th}\sim
10^{12}~\mathrm{s^{-1}}$. Soon after the threshold, the dependence
$N_n(g)$ becomes linear, which means that every quantum of excitation
added to the active medium with a high probability is stimulated to be
emitted as a SP, adding to the coherent SP population. 
%This is a
%consequence of three properties: the pumping rate $g$ dominates over the
%spontaneous decay rate $\gamma_2$, the stimulated SP emission rate is
%much higher than $\gamma_2$, and the population inversion is pinned to a
%constant level from the onset of the spasing [see panel (b) and its the
%discussion below]. 

While this is similar to
conventional lasers, there is a dramatic difference for the spaser. In the
lasers a similar relative rate of the stimulated emission is achieved at
a photon population of $\sim 10^{18}-10^{20}$, while in the spaser the
SP population is $N_n\lesssim 100$. This is due to the much stronger
feedback in spasers because of the much smaller modal volume
$V_n$ -- see discussion of Eq.\
(\ref{qualitative}). The shape of the spasing curves of Fig.\
\ref{Spaser_Threshold_Linewidth_Spectra.eps} (a) (the well-pronounced
threshold with the linear dependence almost immediately above
the threshold) is in an excellent qualitative agreement with the
experiment \cite{Noginov_et_al_Nature_2009_Spaser_Observation}. Note that
the recent SPP nanolaser with a tight transverse confinement of the
lasing modes does not exhibit a pronounced
threshold
\cite{Oulton_Sorger_Zentgraf_Ma_Gladden_Dai_Bartal_Zhang_Nature_2009_Nanolaser},
in contrast to the spaser.

The population inversion number $n_{21}$ as a function of the excitation
rate $g$ is displayed in Fig.\
\ref{Spaser_Threshold_Linewidth_Spectra.eps} (b) for the same set of
frequencies (and with the same color coding) as in panel (a). Before the
spasing threshold, $n_{21}$ increases with $g$ to become positive
with the onset of the population inversion just before the spasing
threshold. For higher $g$,  after the spasing threshold
is exceeded, the inversion  $n_{21}$ becomes constant (the inversion
pinning). 
%Note that both the threshold and pinning level depend on
%frequency, as the SP population does [cf.\ panel (a)]. 
The pinned levels
of the inversion are very low, $n_{21}\sim 0.01$, which again is due to the
very strong feedback in the spaser.

The spectral width $\Gamma_s$ of the spaser generation is due to the
phase diffusion of the quantum SP state caused by the noise of the
spontaneous emission of the SPs into the spasing mode, as described by
Eq.\ (\ref{gamma_s}). This width is displayed in Fig.\
\ref{Spaser_Threshold_Linewidth_Spectra.eps} (c) as a function of the
pumping rate $g$. At the threshold, $\Gamma_s$ is that of the SP line
$\gamma_n$ but for stronger pumping, as the SPs accumulate in the spasing
mode, it decreases $\propto N_n^{-1}$, as given by Eq.\ (\ref{gamma_s}).
This decrease of $\Gamma_s$ reflects the higher coherence of the spasing
state with the increased number of SP quanta and, correspondingly, lower
quantum fluctuations. This is similar to the
lasers as described by the Schawlow-Townes theory
\cite{Schawlow_Townes_PR_1958_Maser_Linewidth}.

%--------------------------------------------------------------------

%--------------------------------------------------------------------

The developed spasing in a dipole SP mode will show itself in the far
field as an anomalously narrow and intense radiation line. The shape and
intensity of this line in relation to the lines of the spontaneous
fluorescence of the isolated gain medium and its SP-enhanced
fluorescence line in the spaser is illustrated in Figs.\
\ref{Spaser_Threshold_Linewidth_Spectra.eps} (d)-(f). Note that for the
system under consideration, there is a 20 meV red shift of the gain
medium fluorescence with respect to the SP line center. It is chosen so
to illustrate the spectral walk-off of the spaser line.
%, which is clearly
%seen in panels (d)-(f). 
For one percent in the
excitation rate above the threshold of the spasing [panel (d)], a broad
spasing line (red color) appears comparable in intensity to
the SP-enhanced spontaneous fluorescence line (blue
color). The width of this spasing line is approximately the same as of
the fluorescence, but its position is shifted appreciably (spectral
walk-off) toward the isolated gain medium line (green color). For the
pumping twice more intense [panel (e)], the
spaser-line radiation dominates, but its width is still close to that of
the SP line due to significant quantum fluctuations of the
spasing state phase. Only when the pumping rate is an order of magnitude
above the threshold, the spaser line strongly narrows [panel (f)], and
it also completely dominates the spectrum of the radiation. This is a
regime of small quantum fluctuations, which is desired in applications.

These results in the spasing region are different in the most
dramatic way from previous phenomenological 
models, which are based on the consideration of 
a gain medium that has negative imaginary part of its
permittivity plus lossy metal nanosystem, described purely
electrodynamically 
\cite{Gordon_Ziolkowski_Ope_Express_2007_Nanoparticle_Laser,%
Wegener_et_al_Opt_Expr_2008_Spaser_Toy_Model}. For instance, in a ``toy
model'' 
\cite{Wegener_et_al_Opt_Expr_2008_Spaser_Toy_Model}, the width of
the resonance line tends to zero at the threshold of spasing and then
broadens up again. This distinction of the present theory
is due the nature of the spasing as
a spontaneous symmetry breaking (nonequilibrium phase transition)
leading to the establishment of the coherent SP state. This state is
influenced by the phase relaxation due to the spontaneous SP emission into
the spasing mode leading to the finite spectral width of the spasing
line $\Gamma_s\propto N_n^{-1}$. Therefore the present theory 
requires a quantum mechanical
consideration of the gain medium. Note that below the spasing threshold,
the phenomenological theories 
\cite{Gordon_Ziolkowski_Ope_Express_2007_Nanoparticle_Laser,%
Wegener_et_al_PRB_2009_Metamaterials_with_Gain} are applicable,
describing such important effects as loss compensation in
metamaterials 
\cite{Shalaev_Nature_Photonics_2007_NIMs}.

%To the contrary, in this theory, the plasmonic 
%fluorescence line
%does not change significantly up to the threshold of spasing, after
%which initially a broad spasing line appears at a generally different
%frequency (a walk-off phenomenon). Only after the spasing threshold is
%significantly exceeded, does the spasing line narrows proportionally to
%the inverse spasing power. Physically this dramatic failure of the toy
%model in the description of the spasing properties stems from the fact
%that the spasing is a result of the instability of the initial state of
%the system, which leads to the spontaneous symmetry breaking. A new state
%is established, which cannot be connected to the initial state by the
%perturbation theory in any order. This new (spasing) state possesses
%a coherent population of the SPs whose phase is defined
%the better the higher is the spasing power. This 
%singularity, which is a non-equilibrium phase transition,
%is fully described by the present theory but completely
%absent in the toy models. 

\subsection{Bistable CW Spaser}
\label{Bistable_Spaser}

Bistables in microelectronics are MOSFET-based devices
that have two stable states describing logical 1 and 0. Their output
``logical'' level changes when the input exceeds a certain threshold value.
Below we show that the spaser can operate as a bistable based on the quantum
amplification.

%--------------------------------------------------------------------
\begin{figure}
\centering
\includegraphics[width=.40\textwidth]
{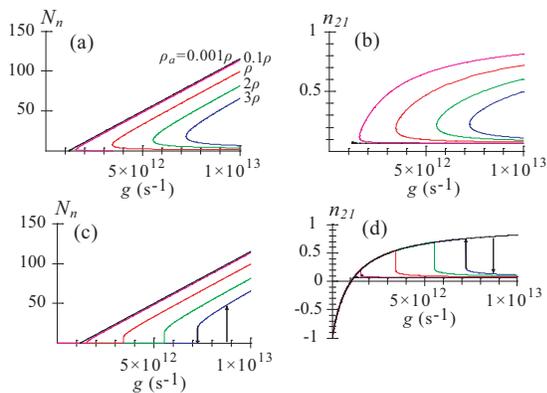}
\caption{
\label{Ag_n2_N_Bistability.eps}
Bistability in spaser with saturable absorber in a stationary spasing
state. (a) Dependence of the SP
population number $N_n$ in the spasing mode on the pumping rate $g$ for
different concentrations $\rho_a$ of the saturable absorber. The curves
are color coded corresponding to $\rho_a$ shown in the units of the
concentration $\rho$ of the active medium chromophores. The black line
shows the threshold curve (separatrix) between the bistable and uniquely 
stable solutions.  (b) The dependence of the population inversion
$n_{21}$ on the pumping rate $g$ for the spasing states. The color
coding is the same as in panel (a). Note that the lower branches of the
curves in this panel correspond to the upper ones in panel (a) and vice
versa. (c) Same as in panel (a) but only the stable branches of the
curves are displayed, illustrating the physical behavior of the
spaser population $N_n$. Note that one of the stable branches coincides
with the  horizontal axis. The vertical arrows illustrate the hysteretic 
transitions between the two stable branches for $\rho_a=3\rho$. 
(d) Same as in panel (c) but for the population inversion $n_{21}$.
The black curve starting with
$n_{21}=-1$ shows a trivial stable solution corresponding to the
absence of spasing ($N_n=0$). 
}
\end{figure}
%--------------------------------------------------------------------

Generally, bistability is a result of nonlinearity in the system. We
examine below the bistability in the spaser where the nonlinearity is
due to the presence of a saturable absorber. Such an absorber is a
chromophore whose absorption overlaps with the spasing line, but which
does not directly absorb the radiation that pumps the spaser. Formally, it is
described by Eqs.\ (\ref{spasing1}), (\ref{n21p_1}) where the pumping
rate $g=0$ for the index $p$ corresponding to the saturable absorber. We
will assume that this absorber is distributed in the space the same way
as the gain chromophores but with a different density $\rho_a$.

Results of a numerical solution of Eqs.\ (\ref{spasing1}),
(\ref{n21p_1}) for different values of $\rho_a$ relatively to the
concentration $\rho$ of the gain medium chromophores are shown in Fig.\
\ref{Ag_n2_N_Bistability.eps}. Panel (a) displays dependence of the SP
population number $N_n$ in the spasing mode on the pumping rate $g$, and
the panel (b) shows the corresponding dependence of the population
inversion $n_{21}$ of the gain medium. We note first that there
always exists also a trivial, non-spasing solution $N_n=0$. For values of $g$
above critical (depending on $\rho_a$), there are also nontrivial
solutions. For $\rho_a>10^{-3}\rho$, these nontrivial solutions consist
of two branches: high-$N_n$ and low-$N_n$. Note that the high-$N_n$
branch in panel (a) corresponds to the low-$n_{21}$ branch in panel (b),
and {\it vice versa}. 

Thus it appears that there is a tri-stability. However, this impression
is incorrect. The lower branches in Fig.\ \ref{Ag_n2_N_Bistability.eps}
(a) [and, correspondingly, the upper branches in panel (b)] describe
unstable solutions that are not realizable physically. This can be
understood already from the fact that along these branches the SP
population decreases with increasing pumping, which is completely
unphysical. 

It is a bistability, not the tri-stability, which takes place in
actuality. This is illustrated in Figs.\
\ref{Ag_n2_N_Bistability.eps} (c) and (d) obtained by isolating the
stable branches of the solutions. As one can see, with an increase of
the pumping rate $g$, the solutions appear at critical pumping rates
that increase with the saturable absorber concentration $\rho_a$. As the
critical $g$ for a given $\rho_a$ is reached, the nonzero-$N_n$ branch
appears with a discontinuity. Both the $N_n=0$ and $N_n>0$ branches are
stable and can keep their states indefinitely. The transition between
these two stable states can be induced by either adding SP quanta to or
removing them from the spasing mode, as illustrated by arrows for
$\rho_a=3\rho$. This shows that a spaser with a saturable
absorber can serve functions of both a nanoscopic memory cell and a
bistable based on quantum amplification. The dynamics of such a device is  
femtosecond as we show below in Sec.\ \ref{Ultrafast}.

In the bistable case there is no perfect pinning of the
population inversion $n_{21}$, as Fig.\ \ref{Ag_n2_N_Bistability.eps}
(d) shows. When the pumping rate $g$ is increased, the system moves
along the stable no-spasing ($N_n=0$) solution denoted by the black
line, where the inversion $n_{21}$ significantly overshoots its values
for the spasing branches (colored lines). When the transition to spasing
occurs, induced, e.g., by an injection of SPs quanta, its is always
discontinuous, as the vertical arrows indicate. Along the spasing
($N_n>0$) branches (colored lines), the inversion counterintuitevely
decreases with increased pumping due to the stimulated emission.

It is instructive to compare the behavior of the spaser shown above in
Fig.\ \ref{Ag_n2_N_Bistability.eps} (see also its discussion) with the
corresponding behavior of lasers. As we have already mentioned in Sec.\
\ref{Introduction}, there are similarities and also substantial differences. A
principal source of these differences is due to the fact the spaser
SP modes are much stronger localized than the photonic modes in
conventional lasers and even the SPP modes in the nanolasers of Refs.\
\onlinecite{Hill_et_al_Opt_Expr_2009_Polaritonic_Nanolaser,%
Oulton_Sorger_Zentgraf_Ma_Gladden_Dai_Bartal_Zhang_Nature_2009_Nanolaser}.
The much smaller modal volume $V_n$ of the spasing SP modes causes
much higher local fields of these SPs [cf.\ Fig.\
\ref{Spaser_Schematic.eps}] and, consequently, much stronger feedback in
the spaser -- cf.\ Eq.\ (\ref{qualitative}). Therefore the spaser
operates, including the bistable operation, at many orders of magnitude
smaller population of SPs (typically, $N_n\sim 100$) per spasing mode as
compared to the number of quanta in the resonators of the conventional
lasers ($\sim 10^{18}-10^{20}$) and even the polaritonic nanolasers.
Related to this fact are the very high values of the local fields in the
spaser $\sim 10^6 \sqrt{N_n}\sim 10^7$ V/cm, as is pointed out above
in this paragraph. These strong local fields
and the strong feedback caused by them are also responsible for the
pronounced linear dependence of the SP population $N_n$ on the pumping rate
[see Fig.\ \ref{Ag_n2_N_Bistability.eps} (c)] and the pronounced
pinning  of the population inversion $n_{21}$ of the gain medium 
[see Fig.\ \ref{Spaser_Threshold_Linewidth_Spectra.eps} (b) and
Fig.\ \ref{Ag_n2_N_Bistability.eps} (d)] in the spasing state.

\section{Spaser as Ultrafast Quantum Nanoamplifier}
\label{Ultrafast}

\subsection{Problem of Setting Spaser as an Amplifier}
\label{Setting}
%--------------------------------------------------------------------
\begin{figure}
\centering
\includegraphics[width=.45\textwidth]
{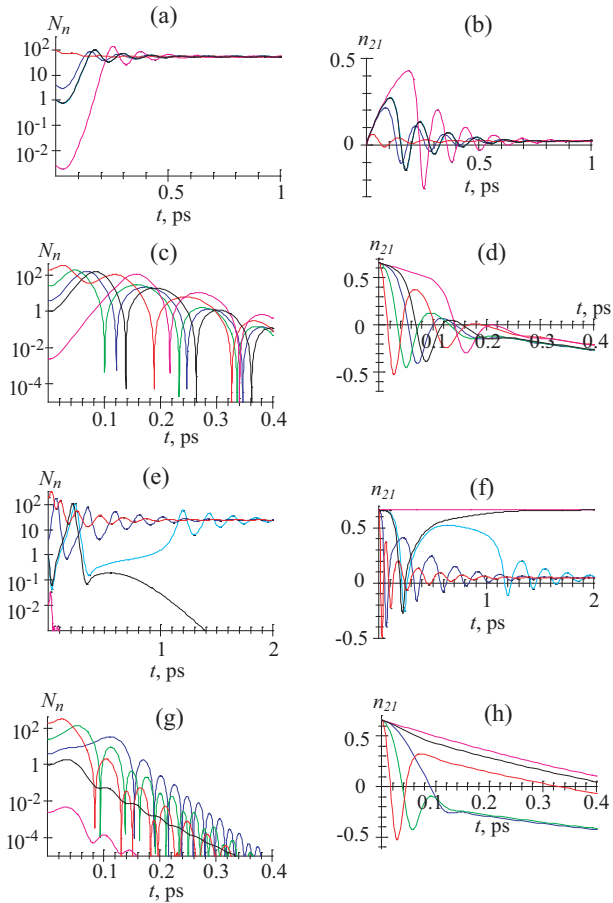}
\caption{\label{Nn_n21_Kinetics.eps}
Ultrafast dynamics of spaser. (a) For monostable spaser (without a
saturable absorber), dependence of SP population in the spasing mode
$N_n$ on time $t$. The spaser is stationary pumped at a rate of
$g=5\times 10^{12}~\mathrm{s^{-1}}$. The color-coded curves correspond
to the initial conditions with the different initial SP populations, as
shown in the graphs. (b) The same as (a) but for the temporal behavior
of the population inversion $n_{21}$. (c) Dynamics of a monostable
spaser (no saturable absorber) with the pulse pumping described as the
initial inversion $n_{21}=0.65$. Coherent SP population $N_n$ is
displayed as a function of time $t$. Different initial populations are
indicated by color-coded curves. (d) The same as (c) but for the
corresponding population inversion  $n_{21}$.
(e) The same as (a) but for bistable
spaser with the saturable absorber in concentration $\rho_a=0.66 \rho$.
(f) The same as (b) but for the bistable spaser. (g) The same as (e) 
but for the pulse pumping with the
initial inversion $n_{21}=0.65$. (h) The same as (g) but for the
corresponding population inversion  $n_{21}$. 
}
\end{figure}

Now we consider the central point of this article: setting the spaser as
an ultrafast quantum nanoamplifier. As we have already mentioned in the
Introduction (Sec.\ \ref{Introduction}), the principal and formidable
problem is that, in contrast to the conventional lasers and amplifiers
in quantum electronics, the spaser has an inherent feedback that cannot
be removed. Thus the spaser will always develop generation and accumulation of
the macroscopic number of coherent SPs in the spasing mode. This leads
to the the population inversion pinning in the CW regime at a very low
level -- cf.\ Fig.\ \ref{Spaser_Threshold_Linewidth_Spectra.eps} (b).
This CW regime corresponds to the net amplification equal zero, which
means that the gain exactly compensates the loss, which condition is
expressed by Eq.\ (\ref{spasing1}). This is a consequence of the
nonlinear gain saturation. This holds for any stable CW generator
(including any spaser or laser) and precludes using them as amplifiers.

We propose two regimes for setting the spaser as a quantum
nanoamplifier. The first is a transient regime based on the fact that
the establishment of the CW regime and the consequent inversion pinning
and total gain vanishing requires some time that is determined mainly by
the rate of the quantum feedback, but depends also on the relaxation
rates of the SPs and the gain medium. After the population inversion is
created by the onset of pumping and before the spasing spontaneously
develops, as we show below in this Section, there is a time interval of
approximately 250 fs, during which the spaser provides usable (and as
predicted, quite high) amplification.

The second way to set the spaser as a quantum nanoamplifier is a bistable
regime that is achieved by introducing a saturable absorber, which
prevents the spontaneous spasing. Then injection of a certain
above-threshold amount of SP quanta will saturate the absorber and
initiate the spasing. Such a bistable quantum amplifier will be
considered in the subsequent Subsection. The bistability is the most
promising regime which allows not only the functioning of the spaser as
a ultrafast threshold (logical) amplifier, but also a quasi-CW bistability
where the spaser works as a memory element [cf. Sec.\
\ref{Bistable_Spaser}] with an ultrafast switching
time.

The temporal behavior of the spaser has been found by direct numerical
solution of Eqs.\ (\ref{rho12})-(\ref{a0n}). This solution 
is facilitated by the fact that in the model under consideration all the
chromophores  experience the same local field inside the
nanoshell, and there are only two types of such chromophores: belonging
to the gain medium and the saturable absorber, if it is present.

\subsection{Monostable Spaser as a Nanoamplifier in Transient Regime}
\label{Transient}

In this Subsection we consider a monostable spaser in a transiebt
regime. This implies that no saturable absorber is present. We
will consider two pumping regimes: stationary and pulse. 

Starting with the stationary regime, we assume that
the pumping at a rate (per one chromophore) of $g=5\times
10^{12}~\mathrm{s^{-1}}$ starts at a moment of time $t=0$ and stays
constant after that. Immediately at $t=0$, a certain number of SPs are
injected into the spaser. We are interested in its temporal dynamics
from this moment on.

The dynamical behavior of the spaser under this pumping regime is
illustrated in Figs.\ \ref{Nn_n21_Kinetics.eps} (a), (b). As we see, the
spaser, which starts from an arbitrary initial population $N_n$, rather
rapidly, within a few hundred femtoseconds approaches the same
stationary (``logical'') level. At this level, an SP population of
$N_n=67$ is established, while the inversion is pinned at a low level of
$n_{21}=0.02$. On the way to this stationary state, the spaser
experiences relaxation oscillations in both the SP numbers and
inversion, which have a trend to oscillate out of phase [compare panels
(a) and (b)]. This temporal dynamics of the spaser is quite complicated
and highly nonlinear (unharmonic). 
It is controlled not by a single relaxation time
but by a set of the relaxation rates. Clearly, among these are the
energy transfer rate from the gain medium to the SPs and the relaxation
rates of the SPs and the chromophores. 

In this mode, the main effect of the initial injection of the SPs
(described theoretically as different initial values of $N_n$) is in the
interval of time it is required for the spaser to reach the final (CW)
state. For very small $N_n$, which in practice can be supplied by the
noise of the spontaneous SP emission into the mode, this time is
approximately 250 fs (cf.: the corresponding SP relaxation time is less
then 50 fs). In contrast, for the initial values of $N_n=1-5$, this time
shortens to 150 fs. Whether this is a practically usable difference
remains to be seen at more advanced stages of development.

Now consider the second regime: pulse pumping. The gain-medium
population of the spaser is inverted at $t=0$ to saturation with a short
(much shorter than 100 fs) pump pulse. Simultaneously, at $t=0$, some
number of plasmons are injected (say, by an external nanoplasmonic
circuitry). In response, the spaser should produce an amplified pulse of
the SP excitation. Such a function of the spaser is illustrated in
Figs.\ \ref{Nn_n21_Kinetics.eps} (c) and (d). 

As we see from panel (c), independently from the initial number of SPs,
the spaser always generates a series of SP pulses, of which only the
first pulse is large (at or above the logical level of $N_n\sim 100$).
(An exception is a case of little practical importance when the initial
$N_n=120$ exceeds this logical level, when two large pulses are
produced.) The underlying mechanism of such a response is the rapid
depletion of the inversion seen in panel (d), where energy is dissipated
in the metal of the spaser. The characteristic duration of the SP pulse
$\sim 100$ fs is defined by this depletion, controlled by the energy
transfer and SP relaxation rates. This time is much shorter than the
spontaneous decay time of the gain medium. This acceleration is due to
the stimulated emission of the SPs into the spasing mode (which can be
called a ``stimulated Purcell effect''). There is also a pronounced trend:
the lower is initial SP population $N_n$, the later the spaser produces
the amplified pulse. In a sense, this spaser functions as a 
pulse-amplitude to time-delay converter.

\subsection{Bistable Spaser (with Saturable Absorber) as an Ultrafast
Nanoamplifier}

Now let us consider the most important point of this article: a bistable
spaser as a quantum threshold (``logical'') nanoamplifier. Such a spaser
contains a saturable absorber mixed with the gain medium with parameters
indicated at the end of Sec.\ \ref{DME}. In particular, the
concentration of the saturable absorber $\rho_a=0.66 \rho$. This case of
a bistable spaser amplifier is of a particular interest because in this
regime the spaser comes as close as possible in its functioning to the
semiconductor-based (mostly, MOSFET-based) digital nanoamplifiers. As in
the previous Subsection, we will consider two cases: the stationary and
short-pulse pumping.

We again start with the case of the stationary pumping at a rate of
$g=5\times 10^{12}~\mathrm{s^{-1}}$. We show in Figs.\
\ref{Nn_n21_Kinetics.eps} (e), (f) the dynamics of such a spaser. For a
small initial population $N_n=5\times 10^{-3}$ simulating the
spontaneous noise, the spaser is rapidly (faster than in 50 fs) relaxing
to the zero population [panel (e)], while its gain-medium population is
equally rapidly approaching a high level [panel (f)] $n_{21}=0.65$
that is defined by the competition of the pumping and the enhanced decay
into the SP mode (the purple curves). This level is so high because the
spasing SP mode population vanishes and the stimulated emission is
absent. After reaching this stable state (which one can call, say,
``logical zero''), the spaser stays in it indefinitely long despite the
continuing pumping.

In contrast, for initial values $N_n$ of the SP population large enough
[for instance, for $N_n=5$, as shown by the blue curves in Figs.\
\ref{Nn_n21_Kinetics.eps} (e) and (f)], the spaser tends to the
``logical one'' state where the stationary SP population reaches the
value of $N_n\approx 60$. Due to the relaxation oscillations, it
actually exceeds this level within a short time of $\lesssim 100$ fs
after the seeding with the initial SPs. As the SP population $N_n$
reaches its stationary (CW) level, the gain medium inversion $n_{21}$ is
pinned down at a low level of a few percent, as typical for the CW
regime of the spaser. This ``logical one'' state salso persists 
indefinitely, as long as the inversion is supported by the pumping.

There is a critical curve (separatrix) that divide the two stable
dynamics types (leading to the logical levels of zero and one). For the
present set of parameters this separatrix starts with the initial
population of $N_n\approx 1$. For a value of the initial $N_n$ slightly
below $1$, the SP population $N_n$ experiences a slow (hundreds
fs in time) relaxation oscillation but eventually relaxes to zero [Fig.\
\ref{Nn_n21_Kinetics.eps} (e), black curve], while the corresponding
chromophore population inversion $n_{21}$ relaxes to the high value
$n_{21}=0.65$ [panel (f), black curve]. In contrast, for a value of
$N_n$ slightly higher than 1 [light blue curves in panels (e) and (f)],
the dynamics is initially close to the separaratrix but eventually the
initial slow dynamics tends to the high SP population and low chromophore
inversion through a series of the relaxation oscillations. The dynamics
close to the separatrix is characterized by a wide range of oscillation
times due to its highly nonlinear character. The initial
dynamics is slowest (the ``decision stage'' of the spaser bistable that
lasts $\gtrsim 1$ ps). The ``decision time'' is diverging
infinitesimally
close to the separatrix, as is characteristic of any threshold (logical)
amplifier.

The gain (amplification coefficient) of the spaser as a threshold
(logical or bistable) amplifier is the ratio of the high CW level to the
threshold level of the SP population $N_n$. For this specific spaser
with the chosen set parameters, this gain is $\approx 60$, which is
more than sufficient for the digital information processing. Thus this
spaser can make a high-gain, $\sim 10$ THz-bandwidth logical amplifier
or dynamical memory cell with excellent prospects of applications. 

The last but probably the most important regime to consider is that of
the pulse pumping in the bistable spaser. In this case, the population
inversion ($n_{21}=0.65$) is created by a short pulse at $t=0$ and
simultaneously initial SP population $N_n$ is created. Both are emulated
as the initial conditions in Eqs.\ (\ref{rho12})-(\ref{a0n}). The
corresponding results are displayed in Figs.\
\ref{Nn_n21_Kinetics.eps} (g) and (h).

When the initial SP population exceeds the critical one of $N_n=1$ (the
blue, green, and red curves), the spaser responds with generating a
short (duration less than 100 fs) pulse of the SP population (and the
corresponding local fields) within a time $\lesssim 100$ fs [panel
(g)]. Simultaneously, the inversion is rapidly (within $\sim 100$ fs)
exhausted [panel (h)].

In contrast, when the initial SP population $N_n$ is less than the
critical one (i.e., $N_n<1$ in this specific case), the spaser rapidly
(within a time $\lesssim 100$ fs) relaxes as $N_n\to 0$ through a series of
realaxation oscillations -- see the black and magenta curves in Fig.\
\ref{Nn_n21_Kinetics.eps} (g). The corresponding inversion
decays in this case almost exponentially with a characteristic time
$\sim 1$ ps determined by the enhanced energy transfer to the SP
mode in the metal -- see the corresponding curves in panel (h). Note
that the SP population decays faster due to the threshold nature of
spasing.

Gold as the spaser nanoplasmonic-core metal, which is used in
experiments \cite{Noginov_et_al_Nature_2009_Spaser_Observation}, has the
relaxation rate an order magnitude higher than silver. Therefore it is
plausible that a gold-core spaser as an amplifier will have a bandwidth
$\sim 100$ THz. We will consider such a system elsewhere. Taking into
account that the local fields produced by the spaser are concentrated
within a few tens of nanometers in its vicinity, these are parameters
that bear promise of many applications in the fundamental science and
coming digital and analog femtosecond technologies.

\section{Discussion and Conclusion}
\label{Conclusions}

We have demonstrated principal possibilities of the spaser functioning
as a ultrafast nanoamplifier of the local optical fields. In doing so we
have had to overcome a principal and formidable problem. Any spaser has
the inherent feedback. Consequently, the spaser as proposed initially
\cite{Bergman_Stockman:2003_PRL_spaser} and observed experimentally
\cite{Noginov_et_al_Nature_2009_Spaser_Observation} rapidly, on the
femtosecond time scale, will generate the coherent population of the SPs
and approach the CW regime. In this regime, the net gain (the saturated
amplification minus loss) is exactly zero (the same as for the lasers).
Thus CW spasers (or CW lasers, for that matter) principally cannot be
used as amplifiers. In the conventional optical amplifiers, in contrast
to lasers, the feedback is deliberately removed and any parasitic
feedback (scattering, etc.) is carefully minimized to prevent the
spontaneous generation. Such a regime is not possible for the
spaser as it is presently known. In this article, we have demonstrated two
ways to set a spaser as a nanoamplifier: (i) the conventional monostable
spaser can amplify in the transient ultrafast (femtosecond) regime
before the CW generation is established; (ii) the spaser with the
saturable absorber can become bistable and function as threshold
(logical) amplifier with ultrafast (femtosecond) switching times.

We have developed quantum theory of the spaser based on density matrix
(optical Bloch) equations. The spaser described is a nanoparticle that
consists of the metal core and the gain (active) medium. The metal core
plays the role of the resonator (``cavity'') whose states are the
localized SPs, which we have quantized. The gain medium is described
completely quantum mechanically taking into account the excitation,
decay, and dephasing processes. The main approximation that we have
employed is that the SP population number $N_n$ is treated as a
classical variable. This is the same semiclassical approximation that is
most often used in the physics of lasers. Given the fact that the
typical number of the SP quanta per spasing mode is still large enough
($N_n\sim 100$), this approximation appears reasonable and reliable. 

Pursuing the goal of the fastest possible operation and the
nanometric size (on the same order as that of microelectronic MOSFETs),
we have deliberately considered spasers whose size is much less than the
radiation wavelength and whose metal thickness is less than the skin
depth. This has allowed us to use the quasistatic SP eigenmodes as full
counterparts of the electrostatic fields in the transistors. Note that
such nanoscopic a spaser has recently been demonstrated experimentally
\cite{Noginov_et_al_Nature_2009_Spaser_Observation}.

%We have supplemented this approximation
%in the spirit of the Schawlow-Townes theory
%\cite{Schawlow_Townes_PR_1958_Maser_Linewidth} by finding the spasing
%modes linewidth $\Gamma_s$ from the quantum fluctuations caused by the
%spontaneous emission of the SPs into the spasing mode. 

Though the main emphasis of this article is on the ultrafast dynamics of
the spaser as a nanoamplifier of the local plasmonic fields, the CW
regime is described first as the basis of the understanding the
ultrafast dynamics. The corresponding results are illustrated in Fig.\
\ref{Spaser_Threshold_Linewidth_Spectra.eps}. The ``spasing curve'' (the
dependence of the SP population $N_n$ on the pumping rate) shows a pronounced
threshold and a linear dependence of the coherent SP population $N_n$ on
the pumping rate $g$ [Fig.\ \ref{Spaser_Threshold_Linewidth_Spectra.eps}
(a)]. This linear dependence $N_n(g)$ is due to the extremely strong
feedback in the spaser caused by the very strong SP mode localization.
The same effect is responsible for clipping the population inversion in
the spasing state to very low levels, as illustrated in Fig.\
\ref{Spaser_Threshold_Linewidth_Spectra.eps} (b). With the increased
pumping, a higher SP population leads to lower quantum fluctuations and
decreased spectral width of the spasing line, as shown in Fig.\
\ref{Spaser_Threshold_Linewidth_Spectra.eps} (c). Gradually, a very
narrow and intense spasing line appears in the emission spetrum of the
spaser -- see Figs.\ \ref{Spaser_Threshold_Linewidth_Spectra.eps}
(d)-(f). This narrow line is indicative of the intense optical fields
that are excited in the vicinity of a spaser -- cf.\ Fig.\
\ref{Spaser_Schematic.eps}(a) where the field must be multiplied by a
factor of $\sqrt{N_n}$. Qualitatively, the linear spasing curve and the
narrowing of the spasing line with the pumping are in a full agreement
with the recent experiment
\cite{Noginov_et_al_Nature_2009_Spaser_Observation}.

Adding a saturable absorber to the spaser shell, we have shown that it
can work as a bistable based on quantum amplification. The corresponding
results are illustrated in Fig.\ \ref{Ag_n2_N_Bistability.eps} for a CW
operation mode.

The main results of this article concern with ultrafast processes of
amplification in the spaser. These are illustrated in Fig.\
\ref{Nn_n21_Kinetics.eps}. We have shown that a monostable spaser
(without a saturable absorber) performs amplification of the injected
SPs in a transient regime. However, this process only manifests itself
in the shorter times to establish the high SP population -- see Figs.\
\ref{Nn_n21_Kinetics.eps} (a)-(d) and the corresponding discussion. In
this sense, such a spaser performs the function of the amplitude to time
converter but not a true nanoamplifier.

The most significant results are those related to the ultrafast
amplification in the bistable spaser containing a saturable absorber
within its gain shell. This amplification has many similarities with
that in semiconductor threshold (logical) nanoamplifiers based on the
MOSFETs. Such a spaser with a stationary pumping behaves as an ultrafast
nanoscale bistable -- see results in Figs.\ \ref{Nn_n21_Kinetics.eps}
(e) and (f). For the initial conditions not too close to the threshold,
it amplifies and switches with a characteristic time of $\sim 100$ fs.
The bistable spaser with the pulse pumping is a high-gain ultrafast
pulse nanoamplifier -- see Figs.\ \ref{Nn_n21_Kinetics.eps} (g) and (h)
and the corresponding discussion in the previous Section. Its typical
response time and pulse length are $\lesssim 100$ fs, which correspond
to the 10 THz bandwidth.

For the gold-based spaser, in contrast to the present
silver-based one, the typical switching time should be much shorter,
presumably $\sim 10$ fs corresponding to $\sim 100$ THz bandwidth.
However, the gold-based spasers will work at a lower SP population due to
the higher loss in the metal. Correspondingly, they will possess lower
signal/noise ratio and wider spectral lines. We will consider such
spasers elsewhere.

Concluding, the spaser is the size of the MOSFET
\cite{Kahng_MOSFET_Patent_1963} ($\sim 10$ nm) and can perform the same
function of signal amplification on the nanoscale. Having a high enough
gain, the spaser can be an active-element foundation of highly
integrated nanoplasmonic devices, including ultrafast processors. Based
on metals, it is inherently much faster, by a factor of $\sim 1000$ than
silicon devices, as we have shown above. For the same reason, it also
highly robust environmentally: it can work at high temperatures, in the
presence of microwave and ionizing radiations, etc. This article has
presented quantum theory of the spaser as a nanoamplifier, which is
critically needed for understanding and utilization of the spaser.

This work was supported by grants from the Chemical Sciences,
Biosciences and Geosciences Division of the Office of Basic Energy
Sciences, Office of Science, U.S. Department of Energy, a grant
CHE-0507147 from NSF, and a grant from the US-Israel BSF. Work at
Garching was supported under contract from Ludwig Maximilian University of
Munich (Germany) in the framework of the Munich  Advanced Photonics Center 
(MAP). I appreciate useful discussions with S.\ Tikhodeev regarding
bistability in lasers.

%% Put the bibliography here, most people will use BiBTeX in
%% which case the environment below should be replaced with
%% the \bibliography{} command.

\bibliography{references}

%% Here is the endmatter stuff: Supplementary Info, etc.
%% Use \item's to separate, default label is "Acknowledgements"

%% Here is the endmatter stuff: Supplementary Info, etc.
%% Use \item's to separate, default label is "Acknowledgements"

%% Put the bibliography here, most people will use BiBTeX in
%% which case the environment below should be replaced with
%% the \bibliography{} command.

\end{document}